 \documentclass[final,3p,times]{elsarticle}
\usepackage{graphics}
\usepackage{graphicx}
\usepackage{amsmath}
\usepackage{amssymb}
\usepackage{amsthm}
\usepackage{color}
\journal{}

\begin{document}

\begin{frontmatter}



%\title{Tunable magnetic and magnetocaloric properties in La$_{0.6}$Ca$_{0.4}$MnO$_3$ nanopowder and nanotubes}
%\title{Dimension dependence on magnetic and Magnetocaloric effect in nanostructured La$_{0.6}$Ca$_{0.4}$MnO$_3$}
\title{Magnetic and magnetocaloric properties of La$_{0.6}$Ca$_{0.4}$MnO$_3$ tunable by particle size and dimensionality}

\author[uff,up]{V. M. Andrade}
\author[uff]{R. J. Caraballo Vivas}
\author[uerj]{S. S. Pedro}
\ead{sandrapedro@uerj.br}
\author[uff]{J. C. G. Tedesco}
\author[cbpf]{A. L. Rossi}
\author[unicamp]{A. A. Coelho}
\author[uff]{D. L. Rocco}
\author[uff]{M. S. Reis}

\address[uff]{Instituto de F\'isica, Universidade Federal Fluminense, 24210-340, Niter\'oi, RJ, Brazil}
\address[up]{Institute of Nanoscience and Nanotechnology, Departamento de F\'isica da Faculdade de Ci\^encias ,Universidade do Porto, 4169-007 Porto, Portugal}
\address[uerj]{Instituto de F\'isica, Universidade do Estado do Rio de Janeiro, 20550-900, Rio de Janeiro, RJ, Brazil}
\address[cbpf]{Centro Brasileiro de Pesquisas F\'isicas, 22290-180, Rio de Janeiro, RJ, Brazil}
\address[unicamp]{Instituto de F\'{i}sica "Gleb Wataghin", Universidade Estadual de Campinas, Caixa Postal 6165, 13083-859, Campinas, SP, Brazil}


\date{\today}

\begin{abstract}
Manganites have been attracted considerable attention due to some intriguing magnetic properties, such as magnetoresistance, spin glass behavior and superparamagnetism. In recent years, some studies point to the effect of particle size and dimensionality of these compounds in their magnetic features. Particularly, LaCaMnO material research is well explored concerning the bulk material. To overcome the lack of the information we successfully produced advanced nanostructures of La$_{0.6}$Ca$_{0.4}$MnO$_{3}$ manganites, namely nanotubes and nanoparticles by using a sol-gel modified method, to determine the size particle effect on the magnetism. The manganites crystal structure, magnetic and magnetocaloric properties were studied in a broad temperature range. Transmission electron microscopy revealed nanoparticles with sizes from 45 up to 223 nm, depending on the calcination temperature. It was found that the magnetic and magnetocaloric properties can be optimized by tuning the particle size; for instance, the magnetic transition broadening by decreasing the particle size. We report the relative cooling power (RCP) of these samples; it was found that the best RCP was observed for the 223 nm particle (508 J/Kg). Finally, this work contributes to the research on the magnetic properties and magnetocaloric potentials in nanostructured systems with distinct morphologies.
\end{abstract}

\end{frontmatter}

% \linenumbers

\section{Introduction}
\label{}



Nowadays, the synthesis of nanomaterials with multifunctional properties is a field of intense activity, especially regarding to technological innovation and development of new materials.
One of the most important motivations to perform the synthesis of nanostructured systems is related to its physical and chemical properties, which may be very distinct from the observed in conventional bulk systems \cite{shankar}. Such differences can be related to surface effects, since nanostructured systems have a considerable surface/volume grain ratio, much higher than the bulk counterpart, causing a large difference in the electronic and magnetic properties \cite{lu}.

The remarkable physical features that these materials exhibit can lead to application fields related to nanomedicine, biomagnetic sensors development, catalysis processes, data storage, logical  and nanoelectronic devices and magnetic refrigeration \cite{kaman,hu,liu,mathew,lee,curiale}, just to mention some examples. In this way, is therefore necessary to discover and to understand the mechanisms that control the observed effects on the physical properties of these materials.

From the point of view of magnetic properties, if the size of the particle has nano dimensions, the material can display
curious and intriguing behaviors, such as spin glass, superparamagnetism, large coercivities and changes in saturation magnetization and Curie temperature \cite{zhu,dey,katiyar,vasilakaki}. The paramagnetic-ferromagnetic phase transition can change from first to second order with the reduction of dimensionality. Consequently, this behavior is directly related to changes in the magnetocaloric effect of the material and the temperature range where the magnetic entropy change occurs tends to become broader with the reduction of the particle size \cite{curiale,xi,pekala,lampen,kumaresavanji,andrade,andrade2}. Such behavior is desirable for the development of magnetic refrigeration systems that can operate in a wide temperature range. From several magnetic materials in nanoscale, manganites are the most promising due to the easy preparation, relatively low cost and remarkable magnetic features \citep{shankar, pekala, andrade, andrade2}.

Based on the features pointed above, the aim of this paper is to investigate the magnetic and magnetocaloric properties of the La$_{0.6}$Ca$_{0.4}$MnO$_3$ nanostructured manganite in distinct morphologies (nanoparticles and nanotubes). The structures were determined using powder X-ray diffraction and the morphology was identified by transmission electron microscopy. From magnetic measurements we evaluated the potential applications of these nanostructures based in their magnetocaloric properties.

\section{Experimental Details}

Sol-gel method (Pechini) was used to prepare nanotubes and nanoparticle samples. It was used analytical grade lanthanum nitrate (La(NO$_3$)$_3$6H$_2$O), calcium carbonate (CaCO$_3$) and manganese acetate (Mn(CH$_3$COO)$_2$4H$_2$O) weight\-ed accurately; the reactants were dissolved into nitric and citric acid solution in deionized water. The obtained mixture was mixed to obtain a clear solution with molar ratio of La:Ca:Mn=0.6:0.4:1. A suitable amount of polyethylene glycol was added to the solution as polymerizing agent. In order to evaporate the excess of solvents and to promote polymerization, the solution was submitted to 343 K for 6 h and then a yellow transparent viscous solution was obtained. The solution was separated in four parts and each part was calcinated at 973 K, 1073 K, 1173 K e 1273 K for 10 h and a final black La$_{0.6}$Ca$_{0.4}$MnO$_3$ powder was obtained for all temperatures.

Nanotubes of the synthesized material were obtained from the pore wetting technique \cite{levy,leyva}: alumina membranes with 200 nm of diameter were immersed in the remained solution early described and kept for 2 h in vacuum. The membranes filled with the material were dried and submitted to calcination for 8 h at 973 K. The final diameters of the nanotubes are determined by the size of the membranes pores used as template.

X-ray powder diffraction data were obtained at UFF at room temperature, using a Bruker AXS D8 Advance diffractometer with Cu-K$\alpha$ radiation ($\lambda$ = 1.54056 \AA{}), 40 kV and 40 mA. Data were collected in the 15$^{o}$ $<$ 2$\theta$ $<$ 85$^{o}$ range in a Bragg-Brentano geometry, with a step size of 0.02$^{o}$ and a counting time of 0.1 s per step. To confirm the formation of the nanotubes, transmission electron microscopy technique (TEM), held at CBPF, was performed using a  Jeol 2100F microscope with an acceleration voltage of 200 kV. For this analysis the samples were diluted in alcohol. Electron tomography was obtained in the TEM to determine the nanotube dimensions. The sample was tilted from $\pm$60$^{o}$ with increments of 1$^{o}$.  Damage related to irradiation was not observed. The alignment and reconstruction of the 3D volume were obtained using \textit{IMOD} software (University of Colorado, USA). \textit{ImageJ} software was used for the visualization of the 3D volume (National Institutes of Health, USA). Magnetic measurements were carried out using a commercial Superconducting Quantum Interference Device (SQUID) at Unicamp.

\section{Crystal structure and morphology}

\begin{figure}[!htb]
\center
\includegraphics[scale=0.35,angle=-90]{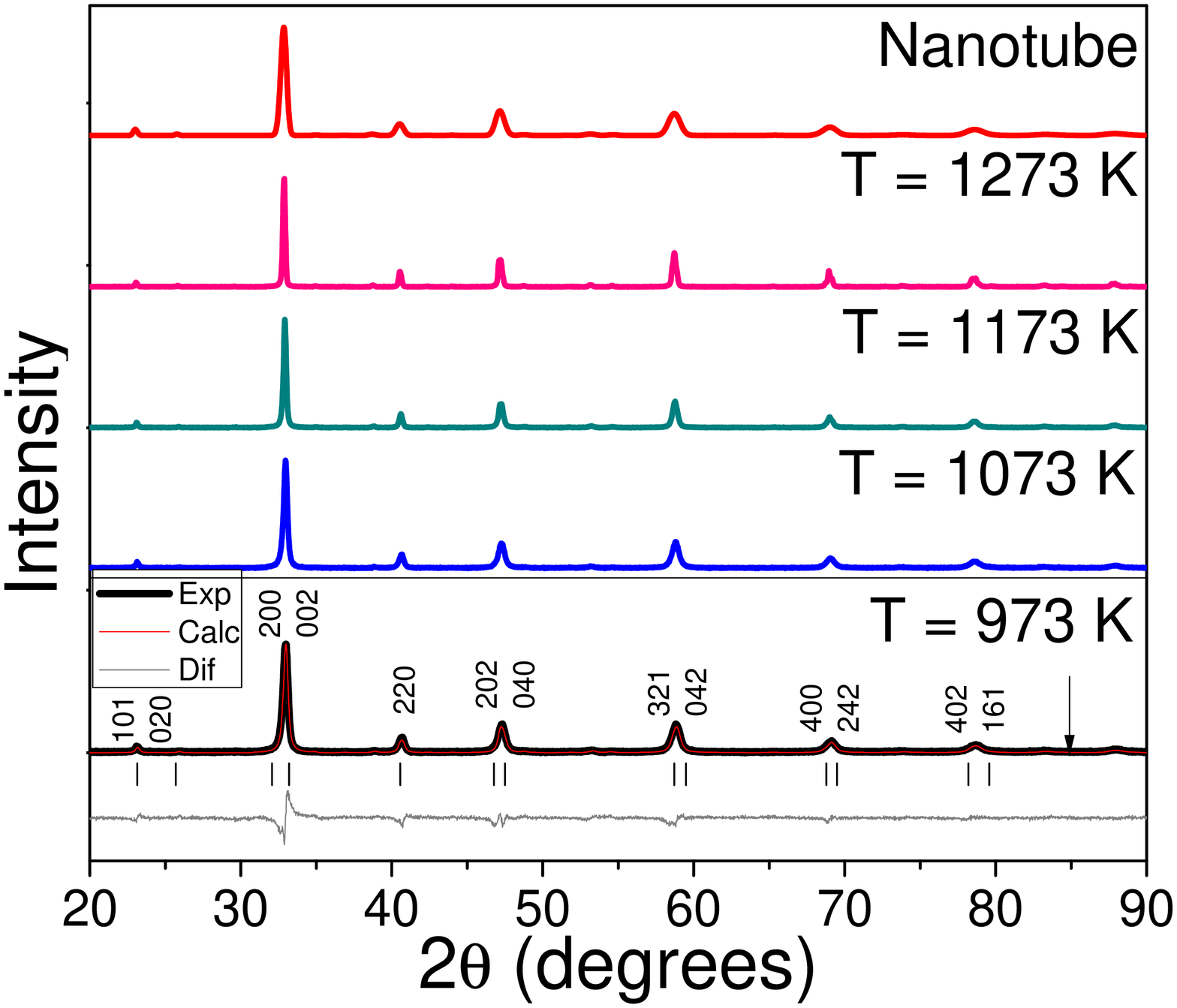}
\caption{Powder X-ray diffraction patterns La$_{0.6}$Ca$_{0.4}$MnO$_3$ nanotubes and nanoparticles. The refined profile of the nanoparticle synthesized at 973 K is shown at the bottom.}
\label{xray}
\end{figure}

X-ray diffractograms of the nanotubes and nanopowders submitted to several calcination temperatures confirmed the formation of a pure La$_{0.6}$Ca$_{0.4}$MnO$_3$ crystalline phase, showing the characteristic peaks of the compound as can be seen in Fig. \ref{xray}. The system presents a perovskite structure at room temperature and pressure, belonging to the space group \textit{Pnma} in an orthorhombic crystal system \cite{sagdeo}. Powder X-ray diffraction data were refined by the Rietveld method. The convergence factors and structure parameters obtained from the Rietveld analysis are summarized in the Table 1 and point to the good quality of the refinement.

\begin{figure*}[!htb]
\centering
\includegraphics[scale=0.9,angle=-90]{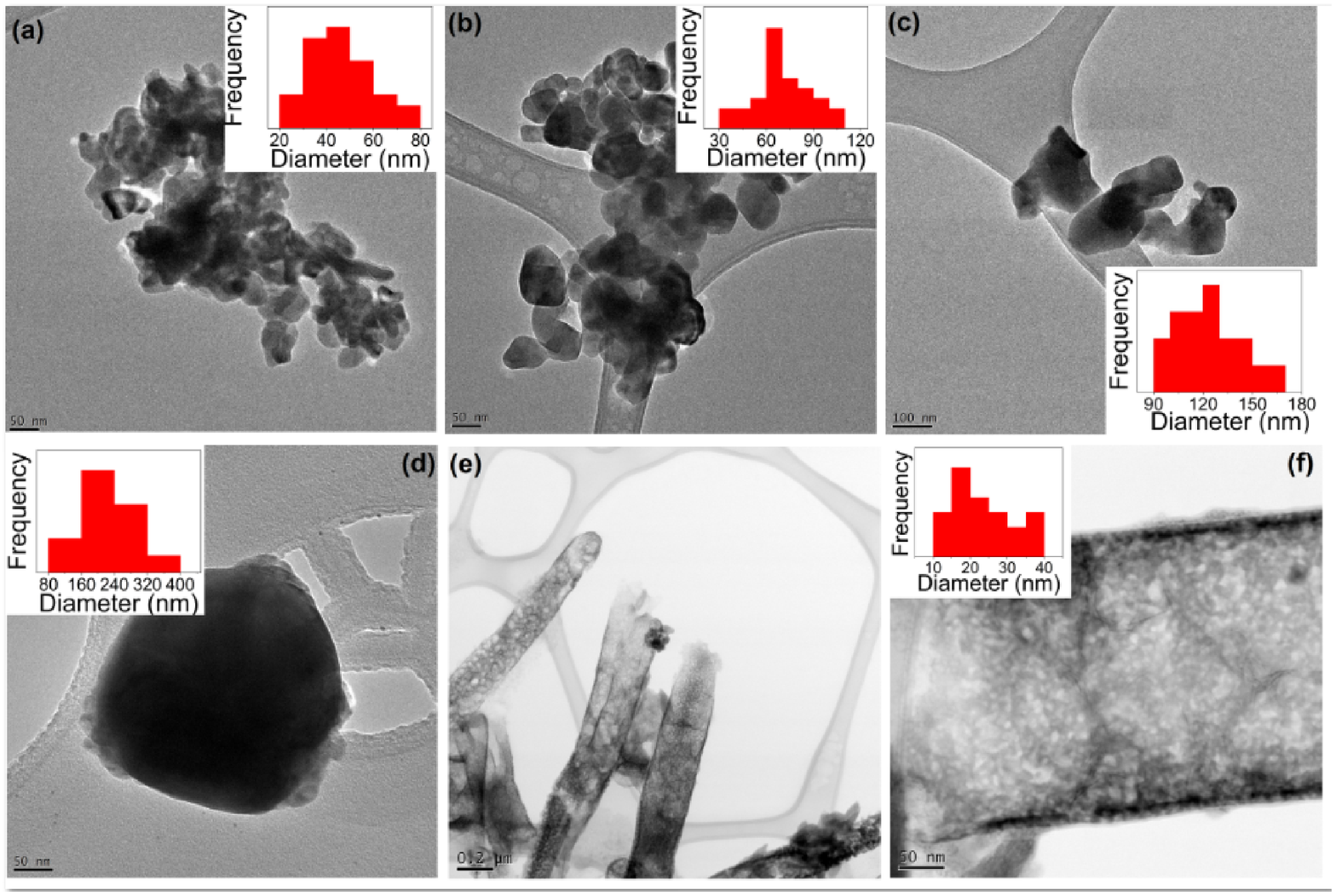}
\caption{TEM images of nanoparticles synthesized at (a) 973 K, (b) 1073 K, (c) 1173 K, (d) 1273 K and (e) nanotubes (at 973 K). In (f) it was obtained an image in high resolution mode to verify the features of the nanoparticles that constitutes the wall of the nanotubes. The histograms for each sample are also shown, where the corresponding values are estimated particle sizes.}
\label{tem}
\end{figure*}

\begin{figure*}[!htb]
\centering
\includegraphics[scale=0.5]{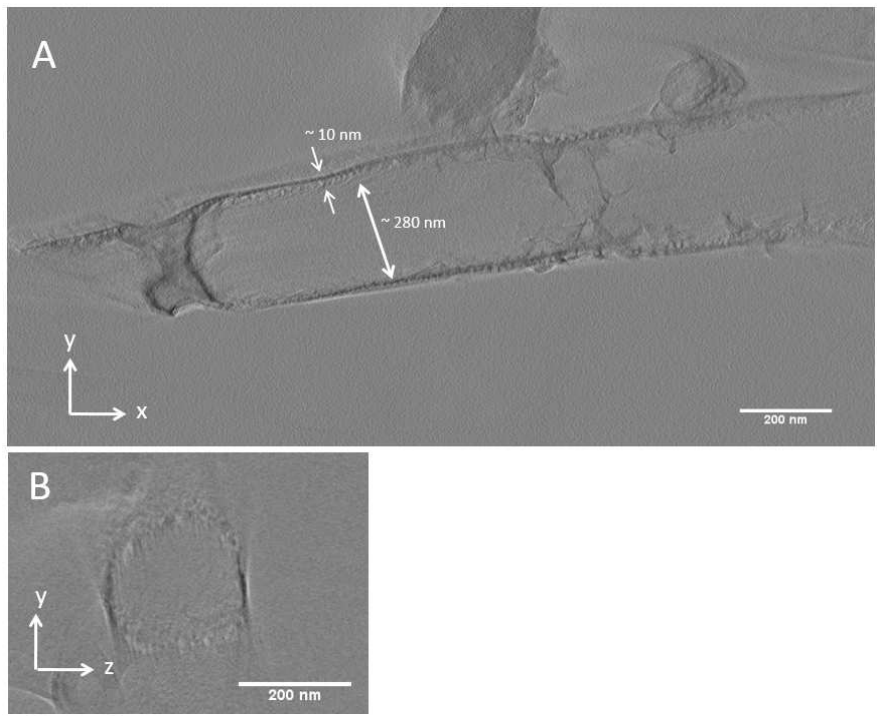}
\caption{Electron tomography and 3D reconstruction of a nanotube. XY (A) and YZ (B) section of the 3D volume. }
\label{tomography}
\end{figure*}

In order to confirm the average grain size, we used Transmission Electron Microscopy (TEM) whose obtained images are shown in Fig. \ref{tem}. We also show in the same figure the histograms of the grain diameter frequencies for each sample, with values obtained from TEM images. The average grain size $D$ on Table 1 indicates that the method was successful in the achievement of nanostructured samples, with a smaller particle size obtained with the smaller calcination temperature, as expected \cite{dey2}. The average grain size increases from 45 nm to 223 nm with the increasing of the calcination temperature for the nanoparticles. TEM images also show that the nanoparticles exhibit a spherical morphology and that are slightly connected to each other. The nanotubes walls are constituted by a set of connected nanoparticles, with average grain size of 23 nm \cite{curiale}. From Fig. \ref{tem}(f) is apparent the granular aspect of the nanotubes formed by nanoparticles. The histograms in the insets of Fig. \ref{tem} exhibit the average grain size for the isolated nanoparticles calcinated at 973, 1073, 1173 and 1273 K and the nanoparticles that constitute the wall of the nanotubes. Electron tomography was obtained in the TEM and are show in the Fig. \ref{tomography} to estimate the diameter of the nanotubes and the wall thickness. The figures show two sections of the 3D volume in the XY (Fig. \ref{tomography}A) and YZ (Fig. \ref{tomography}B) plane. From these images the wall thickness (10 nm) and diameter (280 nm) of the nanotube were estimated.

\newpage

\begin{table*}
\centering
\caption{Refined crystallographic data, reliability factors and average grain size $D$ for La$_{0.6}$Ca$_{0.4}$MnO$_{3}$ nanoparticles.\label{crystaldata}}
    \begin{tabular}{|c|c|c|c|c|c|c|}
     \multicolumn{2}{l}{Parameter} & \multicolumn{2}{l}{Samples}\\\hline
                        & 973 K      & 1073 K     & 1173 K     & 1273 K     & Nanotubes \\\hline\hline
    $a$ (\AA{})         & 5.4669     & 5.4657     & 5.4594     & 5.4542     & 5.4845   \\
    $b$ (\AA{})         & 7.6716     & 7.6800     & 7.6826     & 7.6802     & 7.7092   \\
    $c$ (\AA{})         & 5.4349     & 5.4412     & 5.4411     & 5.4396     & 5.4225   \\
    R$_{p}$ (\%)        & 3.61       & 3.80       & 2.76       & 4.20   	& 2.01     \\
    R$_{wp}$ (\%)       & 4.78       & 4.87       & 3.61       & 5.52       & 2.55     \\
    R$_{exp}$ (\%)      & 2.79       & 3.51       & 2.56       & 2.89       & 0.93     \\
    $\chi^{2}$              & 1.71       & 1.39       & 1.41       & 1.91       & 2.74     \\\hline\hline
    $D (nm)$            &$45 \pm 14$ &$70 \pm 18$ &$122\pm20$  &$223\pm78$  &$23\pm8$   \\\hline
    \end{tabular}
\end{table*}

\section{Magnetic and Magnetocaloric properties}

\begin{figure}[htb!]
\center
\includegraphics[scale=0.4]{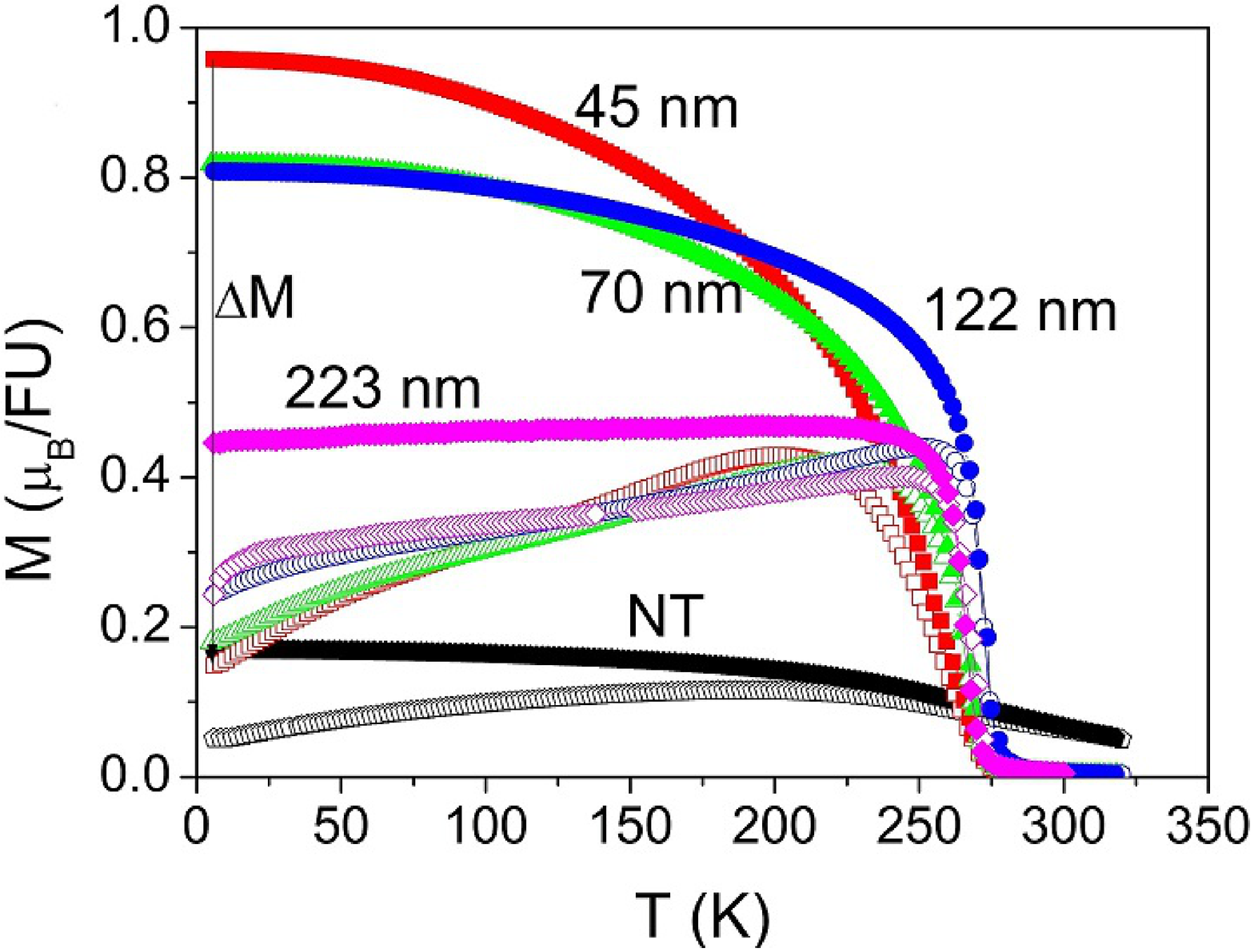}
\caption{Magnetization ZFC and FC as function of temperature at 200 Oe for all samples.}
\label{mvst}
\end{figure}

\begin{figure}[htb!]
\center
\includegraphics[scale=0.4]{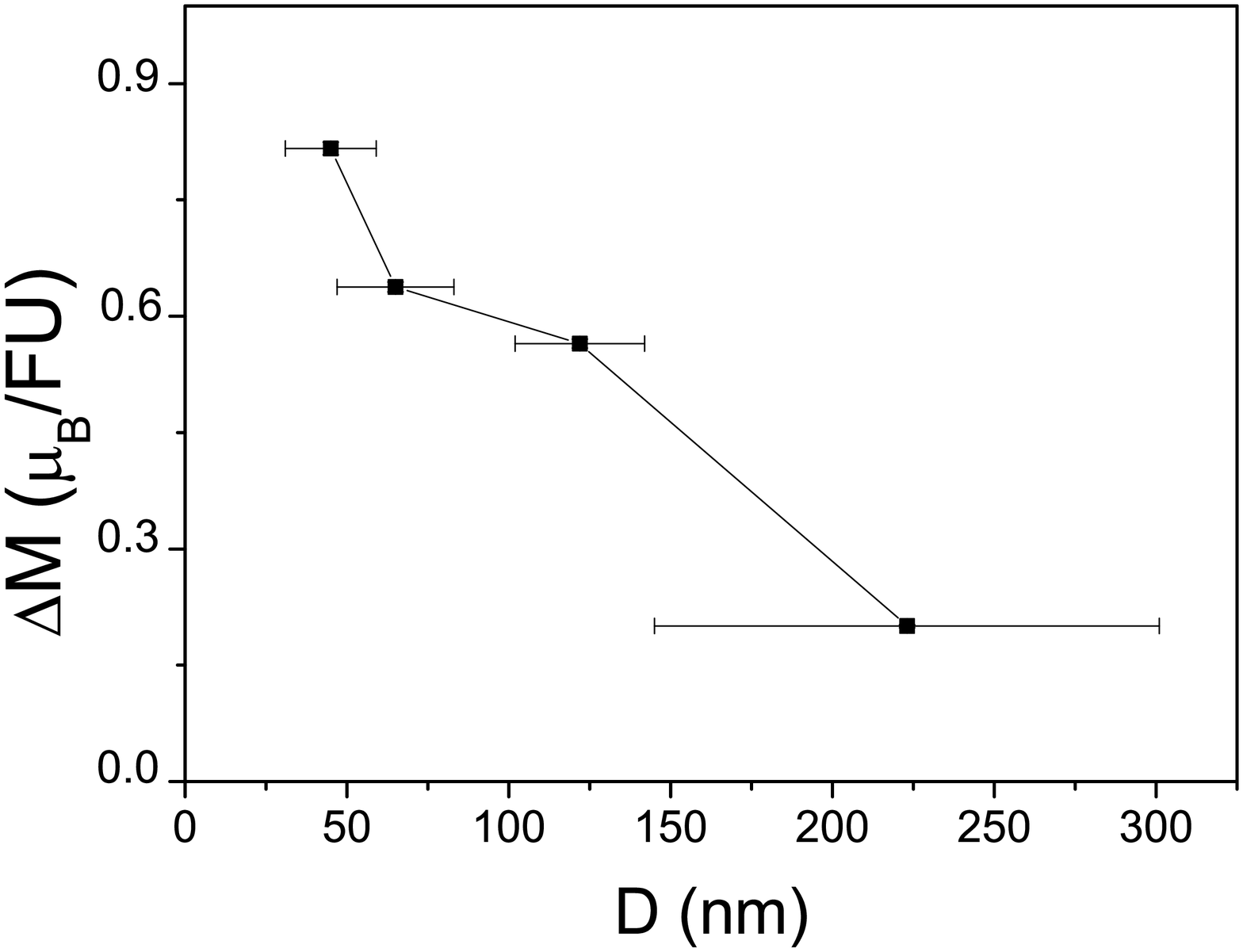}
\caption{Magnetization change between ZFC and FC regimes at 4 K as a function of the particle diameter D.}
\label{d}
\end{figure}

Fig. \ref{mvst} presents the magnetization curves as a function of temperature at zero field cooled (ZFC) and field cooled (FC) under an external magnetic field of 200 Oe, for all samples. Due to the distinct morphology of the samples (nanopowder and nanotubes), it is expected distinct magnetic behaviors for each sample, as observed. We found a large irreversibility $\Delta M$ at 4 K between the ZFC and the FC magnetization for all samples. Fig. \ref{d} shows the $\Delta M$ change with maximum value for the 45 nm nanoparticle and decreases as the nanoparticles size increases. The nanotubes (not shown in Fig. \ref{d}) presented the lowest irreversibility in the $M(T)$ curves. However, it is not possible to make an accurate comparison between $\Delta M$ values for the nanotubes and the nanoparticles, since that besides the nanoparticles forming the nanotubes presented the lowest size, they are grouped in a distinct way of the other powder nanoparticles. In this way, the nanoparticles interact in distinct ways when distributed in powder and when assembled to form the nanotubes, with distinct mechanisms and anisotropy effects regulating the magnetic features of both systems, showing a strong correlation between the structure dimensionality and magnetic features \citep{curiale,curiale3}.

The magnetization curves are strongly dependent on the particle size. For the bulk sample, it is known that the magnetic transition temperature is reported around $T_C$ = 260 K, with $M(T)$ presenting an almost temperature independent behavior up to $T_C$; and above this value it drops to zero \cite{nasri,bohigas}. Concerning the nanoparticles, a sharp ferro-paramagnetic transition is observed for the sample with 223 nm, but decreasing the nanoparticle size this transition broadens. At room temperature, the powder nanoparticles show paramagnetic behavior and exhibit a ferromagnetic transition at Curie temperature ($T_C$) at 258 K (45 nm), 269 K (70 nm), 272 K (122 nm) and 270 K (223 nm), where $T_C$ was defined as the temperature where the derivative $dM/dT$ reaches a minimum value. The reduction in the Curie temperature with decreasing of particle size is usually attributed to finite size effects, due to atomic defects at the surface, leading to disorder in nanocrystalline systems \citep{lampen}.

Below $T_C$, the magnetization increases in a different fashion, depending on particle size. These curves have a hyperbolic-like shape, as already observed in La$_{0.6}$Sr$_{0.4}$MnO$_3$ nanoparticles \cite{andrade}. Concerning the nanotubes, it was not possible to see the magnetic transition in the investigated temperature range. For all cases, the maximum value of magnetization is lower than the complete magnetization of the bulk material. This behaviour was observed for several manganites nanoparticles and can be explained by the presence of the well-known phenomenon of the magnetic dead layer \cite{curiale,lampen,curiale2,ehsani}, where the spins are practically paramagnetic, and located in the nanoparticle shell. The core behaves as the bulk material, whereas the shell behaves like a disordered magnetic system, whose magnetization may be considered to be zero in the absence of magnetic field, as in ordinary paramagnetic systems \cite{ehsani}. The nanoparticle is composed of two different parts: a core, where double exchange interaction dominates and promotes ferromagnetic behavior; and an external part (shell), composed by a layer where magnetic interactions are clearly modified by defects, vacancies, stress, and broke bonds, directing to a disordered magnetic state. In this way, Hueso and co-authors \cite{hueso}, for example, argues that the inner core always retains the intrinsic first-order magnetic transition of the bulk compound, while the disordered shell is more likely to undergo a second-order transition, from the disordered state into the paramagnetic. The composition of the magnetic features of core and shell hides the presence of the first-order magnetic transition. In this way, the overall result in the smallest particles is a broadened second-order transition, although both contribution should be present at the same time. Other authors, however, claim that the magnetization decreasing  in nanosystems can be explained by a simple model where the nanoparticle contains a ferromagnetic core wrapped by a ferrimagnetic shell \cite{andrade}.

\begin{figure}[htb!]
\center
\includegraphics[scale=0.55]{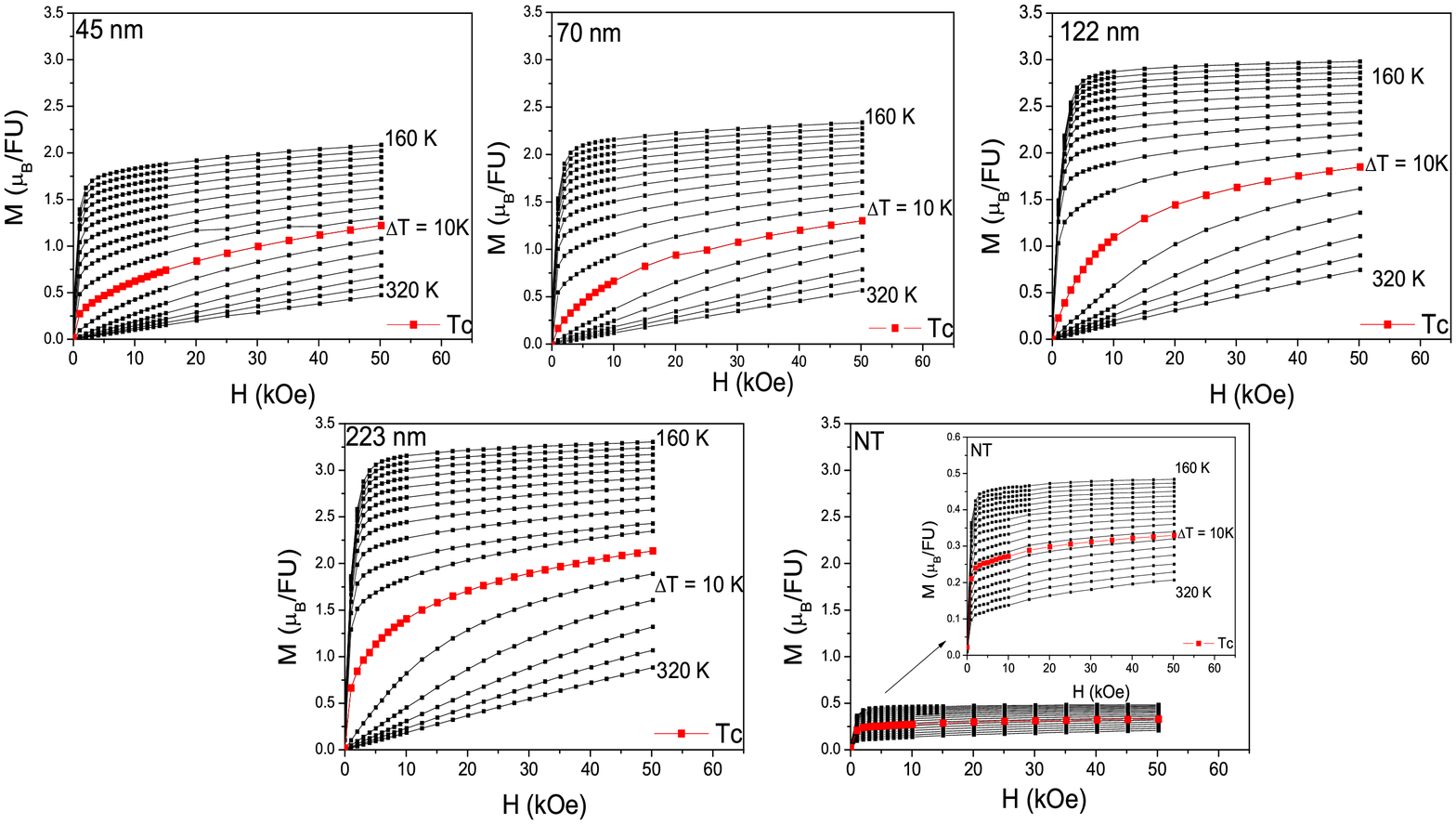}
\caption{Magnetization isotherms for magnetic fields up to 50 kOe. The isotherms corresponding to $T_C$ are highlighted in red for each sample.}
\label{mvsh}
\end{figure}

Magnetization isotherms in the 160-320 K temperature range with $\Delta T$ = 10 K at magnetic fields up to 50 kOe are shown on Figure \ref{mvsh}. The isotherms corresponding to $T_C$ are highlighted in red for each sample. As can be seen, the isotherms present a very similar shape for all samples, with magnetization increasing with nanoparticle size and very close to reach the saturation at 160 K. The isotherms for the nanotubes present a very small value of magnetization when compared to the nanoparticles magnetization.

The physical quantities that measure the magnetocaloric potential are the magnetic entropy change $\Delta S$ and the adiabatic temperature change $\Delta T$. The former can be evaluated by using the measurement of initial isothermal magnetization versus magnetic field at several temperatures, while the latter is measured by the adiabatic change of temperature under the application of a magnetic field. Most commonly, the first method is preferred to avoid the difficulties of adiabatic measurements. In this way, magnetic entropy change is more common to be found in the literature, since it is only needed to know the magnetization map, i.e. $M(T,H)$ to obtain $\Delta S$. Using $M$ \textit{vs.} $H$ isothermal curves it is possible to calculate $\Delta S$ accordingly to \cite{pecharsky}:

\begin{equation}
 \Delta S(T,\Delta H)=\int^{H}_{0}\left(\frac{\partial M(T,H)}{\partial T}\right)_{H}dH.
\label{equ1}
\end{equation}

\begin{figure}[htb!]
\center
\includegraphics[scale=0.55]{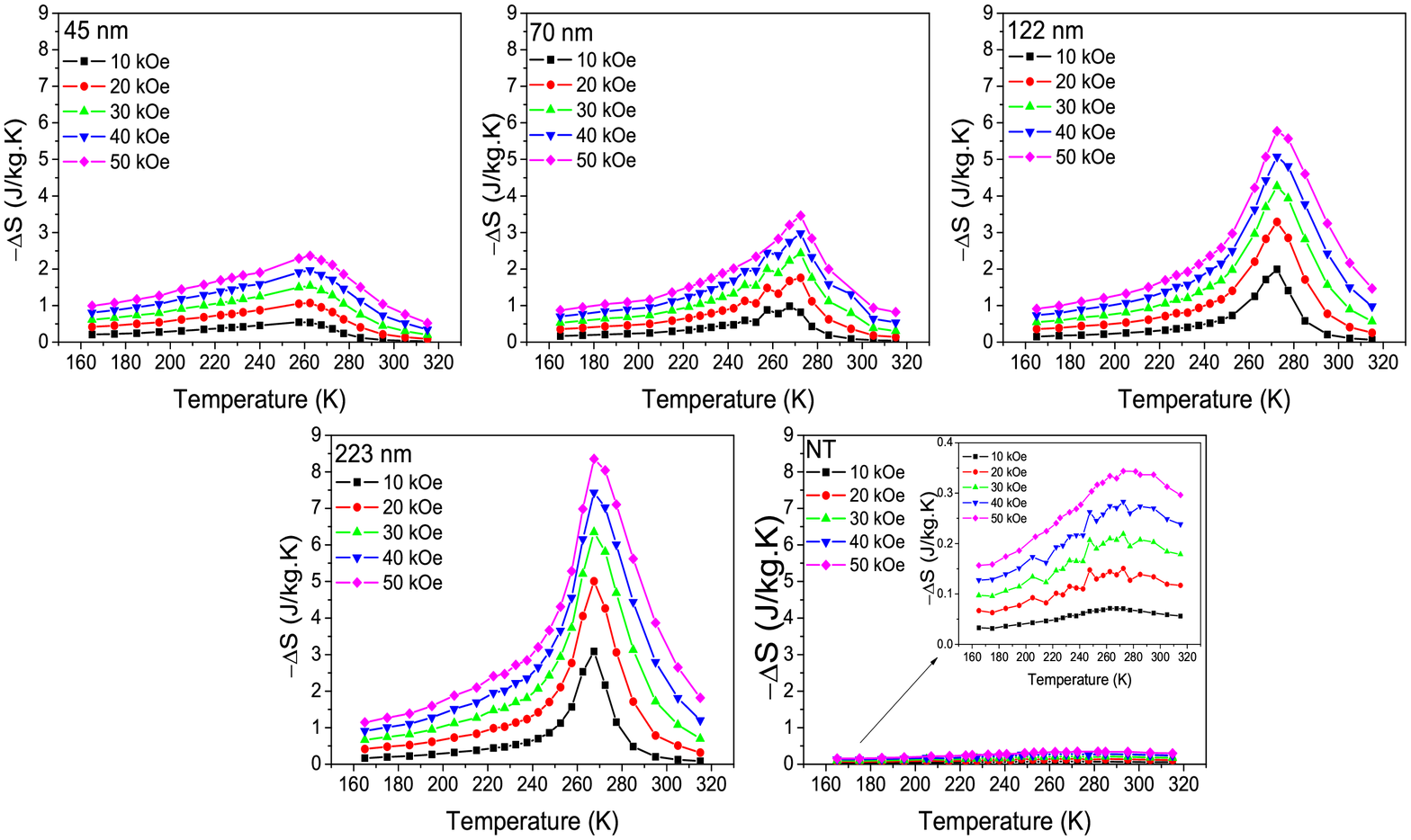}
\caption{Entropy change $\Delta S$ for several magnetic fields up to 50 kOe around Curie temperature for all samples.}
\label{ds}
\end{figure}

The magnetization isotherms around $T_C$ were used to determine the magnetocaloric effect from the entropy change $\Delta S$. The calculated values of $\Delta S$ for $\Delta H$ = 10, 20, 30, 40 and 50 kOe are shown in Fig. \ref{ds}. It is possible to note broad $\Delta S$ curves with peaks around Curie temperature, which typically occurs in systems undergoing second-order phase transitions \citep{wang}. As can be seen from these new figures, the magnetic entropy increase with magnetic field increasing for all samples, with the curves becoming narrower around $T_C$ with nanoparticle size increasing. Lu and Pekala suggests that the decay of MCE with particle size is related to the increasing surface/volume ratio in nanoparticles enhancing structural and magnetic disorder and reflecting in the decrease in magnetic order fluctuation around the ferro-paramagnetic transition \citep{lu,pekala}. The magnetic entropy change presents a maximum value of 8.3 J/kg.K around $T_C$ for the 223 nm particle, and tends to decrease by decreasing the particle size. The reduction in the magnetic interactions and in the maximum entropy change may be attributed to several effects related to surface, size, formation of defects, increment of cationic vacancies, and broken bonds on the particles surface. However, it is also important to notice that $\Delta S$ decreasing is commonly accompanied by a broadening in the $\Delta S$ curve. To make a reasonable comparison between the results in the present work and the literature, the table \ref{magdata} shows the magnetic features of nanoparticles and nanotubes formed by LaCa manganite nanoparticles.

\begin{table*}
\centering
\caption{Particle Diameter (D), magnetic transition temperature ($T_C$), maximum entropy change $\Delta S_{max}$, and relative cooling power (RCP) or refrigerant capacity* (RC) of nanopowder (NP) and nanotubes (NT) formed by LaCa manganites nanoparticles. Bulk La$_{0.6}$Ca$_{0.4}$MnO$_3$ and Gadolinium were included for sake of comparison. \label{magdata}}
    \begin{tabular}{|c|c|c|c|c|c|c|}
%     \multicolumn{2}{l}{Parameter} & \multicolumn{2}{l}{Samples}\\
    \hline
    Sample                               & D (nm)     & $T_C$ (K)  & $\Delta S_{max}$ (J/kg.K) & RCP (J/kg)  & Reference \\\hline\hline
    La$_{0.5}$Ca$_{0.5}$MnO$_3$ (NP)     & 8.3        & 245        & 0.75 (2 T)                & 93          & \cite{pekala}   \\
    La$_{0.6}$Ca$_{0.4}$MnO$_3$ (NT)     & 23         & -          & 0.3   (5 T)               & 40          & this work   \\
    La$_{0.6}$Ca$_{0.4}$MnO$_3$ (NP)     & 45         & 258        & 2.3  (5 T)                & 228         & this work   \\
    La$_{0.6}$Ca$_{0.4}$MnO$_3$ (NP)     & 70         & 269        & 3.5  (5 T)                & 251         & this work   \\
    La$_{0.6}$Ca$_{0.4}$MnO$_3$ (NP)     & 122        & 272        & 5.8  (5 T)                & 374         & this work   \\
    La$_{0.6}$Ca$_{0.4}$MnO$_3$ (NP)     & 223        & 270        & 8.3  (5 T)                & 508         & this work   \\
    La$_{0.6}$Ca$_{0.4}$MnO$_3$ (bulk)   & -          & 264        & 5.5  (5 T)                & 139         & \cite{nasri}   \\
    La$_{0.67}$Ca$_{0.33}$MnO$_3$ (NT)   & 28         & 258        & -                         & -           & \cite{curiale}  \\
    La$_{0.7}$Ca$_{0.3}$MnO$_3$ (NP)     & 15         & 241        & 2.6 (5 T)                 & 160*        & \cite{lampen}   \\
    La$_{0.7}$Ca$_{0.3}$MnO$_3$ (NP)     & 33         & 260        & 4.9 (5 T)                 & 150*        & \cite{lampen} \\
    La$_{0.7}$Ca$_{0.3}$MnO$_3$ (NT)     & -          & 236        & 1.9 (5 T)                 & 108*        & \cite{kumaresavanji} \\
    La$_{0.8}$Ca$_{0.2}$MnO$_3$          & 17         & 234        & 0.6 (4.5 T)               & 150         & \cite{xi}   \\
    La$_{0.8}$Ca$_{0.2}$MnO$_3$          & 28         & 214        & 4.5 (4.5 T)               & 350         & \cite{xi}   \\
    La$_{0.8}$Ca$_{0.2}$MnO$_3$          & 43         & 236        & 8.6 (4.5 T)               & 200         & \cite{xi}   \\
    Gd (bulk)                            & -          & 293        & 9.4 (5 T)                 & 690         & \cite{wang}   \\ \hline%\hline
    \end{tabular}
\end{table*}

\begin{figure}[htb!]
\center
\includegraphics[scale=0.5]{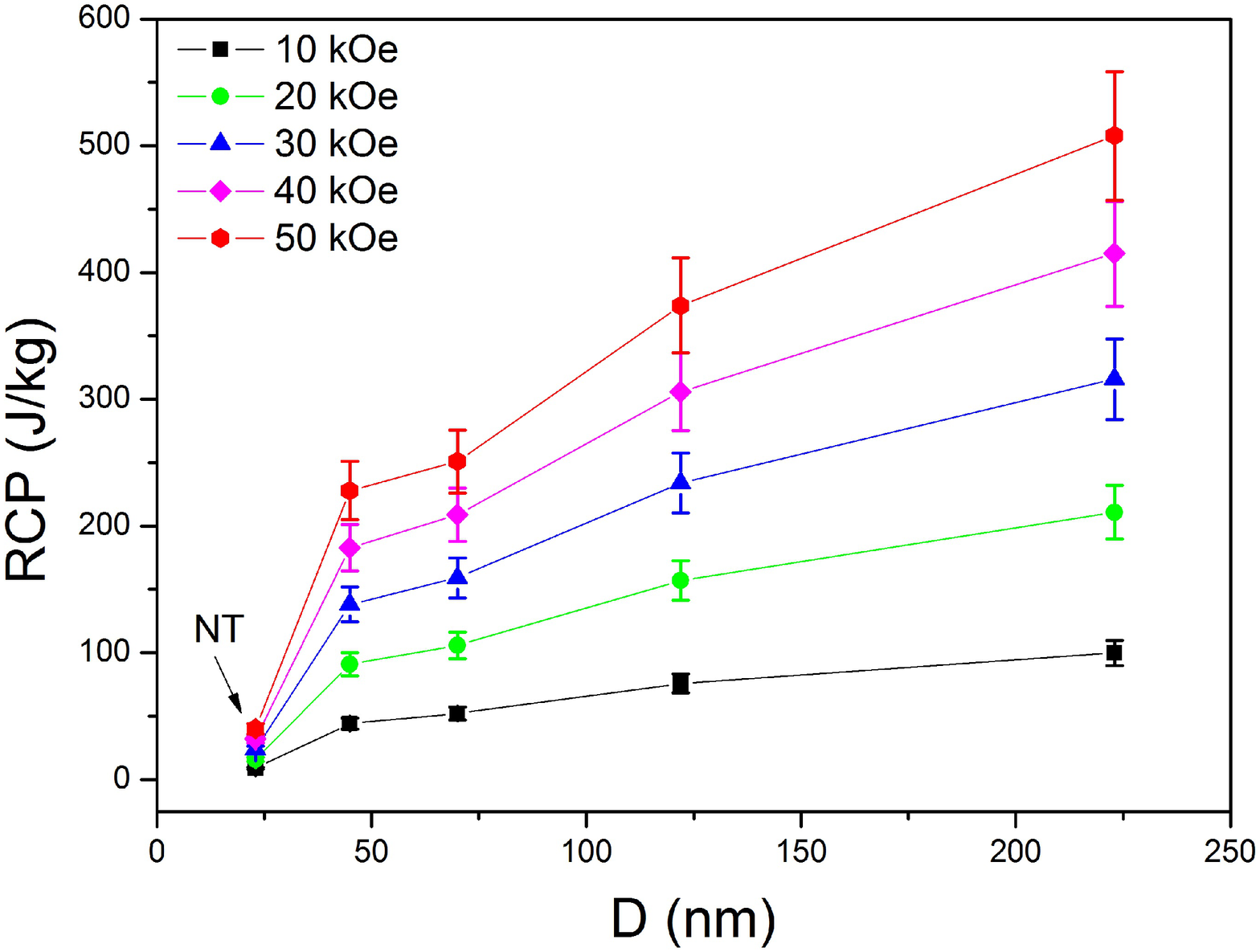}
\caption{Relative Cooling Power (RCP) as a function of particle size $D$ for several magnetic fields up to 50 kOe. Horizontal error bars were omitted for clarity.}
\label{rcp}
\end{figure}

From the results of the magnetic entropy change, it was determined the relative cooling power (RCP), a parameter used to evaluate the refrigeration capacity of a magnetic refrigerant \cite{wang}. The RCP value is obtained from

\begin{equation}
 RCP = |\Delta S_{T,max}|\times \delta T_{FWHM}
\label{equ2}
\end{equation}

\noindent where $|\Delta S_{T,max}|$ is the absolute value of the maximum magnetic entropy change and  $\delta T_{FWHM}=T_2-T_1$  is the temperature difference at the full width at half maximum (FWHM). RCP measures the heat amount to be transferred between $T_1$ and $T_2$ by a magnetic refrigerant material in an ideal cycle \cite{wang}. The calculated values can be seen as a function of $D$  for several magnetic fields up to 50 kOe in the Fig. \ref{rcp}. It is possible to notice that this value changes from 508 J/kg for the 223 nm particle to 40 J/kg for the nanotubes at 50 kOe, which indicates that the RCP of nanoparticles decreases by decreasing the particle size, but they still possess a larger cooling power than the nanotubes of the same compound, due to the broadening of the magnetic transition observed in these samples. In this way, it is important to notice that the reduced maximum value of $\Delta S$ observed for nanosystems is often accompanied by a broad magnetic entropy change.

\section{Conclusions}

Nanoparticles and nanotubes of La$_{0.6}$Ca$_{0.4}$MnO$_3$ manganites were produced by a modified sol-gel method. Structure and morphologies were investigated by powder DRX and TEM techniques, which showed that the nanoparticles change their size with calcination temperature from an average size of 45 nm at 973 K to 223 nm at 1273 K and the nanoparticles assembled to form the nanotubes present an average diameter of 23 nm. Magnetization measurements revealed a hyperbolic-like behavior for these nanostructures, with magnetic transitions between 258 and 272 K, tunable accordingly nanoparticle size. The ferro-paramagnetic phase transition is  modified from first order, as seen in bulk systems, to second order to nanoparticles; but the sample with larger size (223 nm) exhibited a regular magnetic behavior, similar to the bulk sample. Consequently, the magnetocaloric effect (MCE) is modified by changes on the magnetic properties of the samples. The observed temperature interval of the magnetic entropy change broadens when the particle size is reduced to nanometers. In this way, the magnetocaloric effect is strongly particle size dependent and optimized at larger sizes. The relative cooling power (RCP), used to evaluate the refrigeration capacity of a magnetic systems, has its best value for the 223 nm particle. Indeed, the RCP values of LaCa manganites are smaller than that of pure Gd metal, however these manganites have a relatively low cost when compared to Gd, with ease of fabrication and a tunable $T_C$ according to chemical composition.

\section{Acknowledgments}
Access to Laborat\'{o}rio de Difra\c{c}\~{a}o de Raios-X at IF - UFF (Niter\'{o}i, Brazil), and Laborat\'{o}rio de Baixas Temperaturas at UNICAMP (Campinas, Brazil) is gratefully acknowledged by all authors. The authors also acknowledge FAPERJ, FAPESP, CAPES, CNPq, and PROPPI-UFF for financial support.

%% References


\bibliographystyle{model1a-num-names}

\begin{thebibliography}{00}



\bibitem{shankar} K. S. Shankar, S. Kar, A. K. Raychaudhuri and G. N. Subbanna, Fabrication of ordered array of nanowires of La$_{0.67}$Ca$_{0.33}$MnO$_3$ (x=0.33) in alumina templates with enhanced ferromagnetic transition temperature, Appl. Phys. Lett. 84 (2004) 993-995.
\bibitem{lu} W. J. Lu, X. Luo, C. Y. Hao, W. H. Song and Y. P. Sun, Magnetocaloric effect and Griffiths-like phase in La$_{0.67}$Sr$_{0.33}$MnO$_3$ nanoparticles J. Appl. Phys. 104 (2008) 113908.
\bibitem{kaman} O. Kaman, E. Pollert, P. Veverka, M. Veverka, E. Hadov\'{a}, K. Kn\'{i}zek, M. Marysko, P. Kaspar, M. Klementov, V. Gr\"{u}nwaldov\'{a}, S. Vasseur, R. Epherre, S. Mornet, G. Goglio and E. Duguet, Silica encapsulated manganese perovskite nanoparticles for magnetically induced hyperthermia without the risk of overheating. Nanotechnology 20 (2009) 275610.
\bibitem{hu} X. J. Hu, J. K. Liu and Y. Mu. Facile synthesis and characterization of hydroxylapatite nanoparticle chains, Mater. Lett. 62 (2008) 3824-3826.
\bibitem{liu} J. Liu, Z. Zhao, J. Wang, C. Xu, A. Duan, G. Jiang and Q. Yang, The highly active catalysts of nanometric CeO2-supported cobalt oxides for soot combustion, Appl. Catal. B 84 (2008) 185-195.
\bibitem{mathew} S. P. Mathew and S. N. Kaul, Tuning magnetocaloric effect with nanocrystallite size, Appl. Phys. Lett. 98 (2011) 172505.
\bibitem{lee} J. Lee, D. Suess, T. Schrefl, Kyu Hwan Ohm and Josef Fidler, Magnetic characteristics of ferromagnetic nanotube, J. Magn. Magn. Mater. 310 (2007) 2445-2447.
\bibitem{curiale} J. Curiale, R. D. Sanchez, H. E. Troiani, C. A. Ramos, H. Pastoriza, A. G. Leyva and P. Levy, Magnetism of manganite nanotubes constituted by assembled nanoparticles, Phys. Rev. B 75 (2007) 224410.
\bibitem{zhu} T. Zhu, B. G. Shen, J. R. Sun, H. W. Zhao and W. S. Zhan, Surface spin-glass behavior in La 2/3 Sr 1/3 MnO 3 nanoparticles, Appl. Phys. Lett. 78 (2001) 3863-3865.
\bibitem{dey} P. Dey, T. K. Nath, P. K. Manna and S. M. Yusuf, Enhanced grain surface effect on magnetic properties of nanometric La0.7Ca0.3MnO3 manganite: Evidence of surface spin freezing of manganite nanoparticles, J. Appl. Phys. 104 (2008) 103907.
\bibitem{katiyar} P. Katiyar, D. Kumar, T. K. Nath, A. V. Kvit, J. Narayan, S. Chattopadhyay, W. M. Gilmore, S. Coleman, C. B. Lee, J. Sankar and R. K. Singh, Magnetic properties of self-assembled nanoscale La 2/3 Ca 1/3 MnO 3 particles in an alumina matrix, Appl. Phys. Lett. 79 (2001) 1327-1329.
\bibitem{vasilakaki} M. Vasilakaki and K. N. Trohidou, Numerical study of the exchange-bias effect in nanoparticles with ferromagnetic core/ ferrimagnetic disordered shell morphology, Phys. Rev. B 79 (2009) 144402.
\bibitem{xi} S. Xi, W. Lu and Y. Sun, Magnetic properties and magnetocaloric effect of La0.8Ca0.2MnO3 nanoparticles tuned by particle size, J. Appl. Phys. 111 (2012) 063922.
\bibitem{pekala} M. Pekala, V. Drozd, J.F. Fagnard and Ph. Vanderbemden, Magnetocaloric effect in nano- and polycrystalline manganites La0.5Ca0.5MnO3, J. All. Comp. 507 (2010) 350-355.
\bibitem{lampen} P. Lampen, N. S. Bingham, M. H. Phan, H. Kim, M. Osofsky, A. Piqu\'{e}, T. L. Phan, S. C. Yu, and H. Srikanth, Impact of reduced dimensionality on the magnetic and magnetocaloric response of La0.7Ca0.3MnO3, Appl. Phys. Lett. 102 (2013) 062414.
\bibitem{kumaresavanji} M. Kumaresavanji, C. T. Sousa, A. Pires, A. M. Pereira, A. M. L. Lopes, and J. P. Araujo, Magnetocaloric effect in La0.7Ca0.3MnO3 nanotube arrays with broad working temperature span, J. Appl. Phys. 117 (2015) 104304.
\bibitem{andrade} V. M. Andrade, R. J. Carballo-Vivas, T. Costa-Soares, S. S. Pedro, D. L. Rocco, M. S. Reis, A. P. C. Campos and A.A. Coelho, Magnetic and structural investigations on La0.6Sr0.4MnO3 nanostructured manganite: Evidence of a ferrimagnetic shell, J. Sol. St. Chem. 219 (2014) 87-92.
\bibitem{andrade2} V. M. Andrade, S. S. Pedro, D. L. Rocco, M. S. Reis, A. P. C. Campos, A. A. Coelho, M. T. Escote and A. Zenatti, Magnetocaloric functional properties of Sm0.6Sr0.4MnO3 manganite due to advanced nanostructured morphology, in submission (2015).
\bibitem{levy} P. Levy, A. G. Leyva, H. E. Troiani and R. D. S\'{a}nchez, Nanotubes of rare-earth manganese oxide, Appl. Phys. Lett. 83 (2003) 5247-5249.
\bibitem{leyva} A. G. Leyva, P.Stoliar, M. Rosenbusch, P. Levy, J. Curiale, H. Troiani and R. D. Sanchez, Synthesis route for obtaining manganese oxide-based nanostructures, Physica B 354 (2004) 158-160.
\bibitem{sagdeo} P. R. Sagdeo, S. Anwar and N. P. Lalla, Powder X­ray diffraction and Rietveld analysis of La1-xCaxMnO3($0<X<1$), Powder Diffraction 21 (2006) 40.
\bibitem{dey2} P. Dey and T.K. Nath, Enhanced grain surface effect on magnetic properties of nanometric La 0.7 Ca 0.3 MnO3 manganite: Evidence of surface spin freezing of manganite nanoparticles, Phys. Rev. B. 73 (2006) 214425.
\bibitem{curiale3} J. Curiale, R.D. S\'{a}nchez, H.E. Troiani, H. Pastoriza, P. Levy and A.G. Leyva, Morphological and magnetic characterization of manganites oxide-based nanowires and nanotubes, Physica B 354 (2004) 98-103.
\bibitem{nasri} M. Nasri, M. Trik, E. Dhahri, M. Hussein, P. Lachkar and E.K. Hlil, Investigation of structural, magnetocaloric and electrical properties of La0.6Ca0.4xSrxMnO3 compounds, Physica B 408 (2013) 104-109.
\bibitem{bohigas} X. Bohigas, J. Tejada, M. L. Mar\'{i}nez-Sarri\'{o}n, S. Tripp and R. Black, Magnetic and calorimetric measurements on the magnetocaloric e!ect in La0.6Ca0.4MnO3, J. Magn. Magn. Mater. 208 (2000) 85-92.
\bibitem{curiale2} J. Curiale, R. S. Sanch\'ez, H. E. Troiani, A. G. Leyva and P. Levy, Magnetic interactions in ferromagnetic manganite nanotubes of different diameters, Appl. Surf. Sci. 254 (2007) 368.
\bibitem{ehsani} M. H. Ehsani, P. Kameli, M. E. Ghazi, F. S. Razavi, and M. Taheri, Tunable magnetic and magnetocaloric properties of La0.6Sr0.4MnO3 nanoparticles, J. Appl. Phys. 114 (2013) 223907.
\bibitem{hueso}  L. E. Hueso, P. Sande, D. R. Miguéns, J. Rivas, F. Rivadulla and M. A. López-Quintela, Tuning of the magnetocaloric effect in La0.67Ca0.33MnO3-$\delta$ nanoparticles synthesized by sol–gel techniques, J. Appl. Phys 91 (2002) 9943.
\bibitem{pecharsky} V. Pecharsky and K. A. Gschneidner Jr., Giant Magnetocaloric Effect in Gd5(Si2Ge2), Phys. Rev. Lett. 78 (1994) 4494.
\bibitem{wang} G.F. Wang, L.R. Li, Z.R. Zhao, X.Q. Yu, X.F. Zhang, Structural and magnetocaloric effect of Ln0.67Sr0.33MnO3 (Ln=La, Pr and Nd) nanoparticles, Cer. Int. 40 (2014) 16449-16454.
\end{thebibliography}



\end{document}